\documentclass[journal,11pt,onecolumn,draftclsnofoot,]{IEEEtran}
\IEEEoverridecommandlockouts
\usepackage{graphicx}                % For including figures
\usepackage{amsmath}                 % For advanced math formatting
\usepackage{amsfonts}                % For additional math fonts
\usepackage{amssymb}                 % For more math symbols
\usepackage{cite}                    % For managing citations
\usepackage{hyperref}                % For hyperlinks in the document
\usepackage{tabularx}                % For flexible table formatting
\usepackage{multirow}                % For multi-row cells
\usepackage{booktabs}                % For professional table lines
\usepackage{float}                   % For [H] placement option
\usepackage{array}
\usepackage{xcolor} % Optional for cell shading
\usepackage{adjustbox} % For table adjustments
\usepackage{wrapfig} % For wrapping figures or tables
\usepackage{cite}
\usepackage{amsmath,amssymb,amsfonts}
\usepackage{algorithmic}
\usepackage{graphicx}
\usepackage{textcomp}
\usepackage{xcolor}
\def\BibTeX{{\rm B\kern-.05em{\sc i\kern-.025em b}\kern-.08em
    T\kern-.1667em\lower.7ex\hbox{E}\kern-.125emX}}

\usepackage{float}

\usepackage{tabularray}
\usepackage{algorithm}
\usepackage{algorithmic}
\usepackage[USenglish]{babel}
\usepackage[nodayofweek,level]{datetime}
\usepackage[utf8]{inputenc}
\usepackage{rotating}

\usepackage{pdflscape}
\usepackage{afterpage}
\usepackage{capt-of}% or use the larger `caption` package

\usepackage{lipsum}% dummy tex
\usepackage{cite}
\usepackage{csquotes}

\usepackage{tabu}
\def\tsc#1{\csdef{#1}{\textsc{\lowercase{#1}}\xspace}}
\tsc{WGM}
\tsc{QE}
\tsc{EP}
\tsc{PMS}
\tsc{BEC}
\tsc{DE}
\usepackage{framed}

\usepackage{bm}

\usepackage{nomencl}
\usepackage{bbm}

\renewcommand\nomgroup[1]{%
    \item[\bfseries
     \ifstrequal{#1}{O}{\textbf{Operators}}{%
        \ifstrequal{#1}{S}{\textbf{Indices and sets}}{%
            \ifstrequal{#1}{V}{\textbf{Variables and parameters}}{%
            }}}
            ]}
            
            \makenomenclature

\makeatletter
\newsavebox\myboxA
\newsavebox\myboxB
\newlength\mylenA

\newcommand*\xoverline[2][0.75]{%
    \sbox{\myboxA}{$\m@th#2$}%
    \setbox\myboxB\null% Phantom box
    \ht\myboxB=\ht\myboxA%
    \dp\myboxB=\dp\myboxA%
    \wd\myboxB=#1\wd\myboxA% Scale phantom
    \sbox\myboxB{$\m@th\overline{\copy\myboxB}$}%  Overlined phantom
    \setlength\mylenA{\the\wd\myboxA}%   calc width diff
    \addtolength\mylenA{-\the\wd\myboxB}%
    \ifdim\wd\myboxB<\wd\myboxA%
      \rlap{\hskip 0.5\mylenA\usebox\myboxB}{\usebox\myboxA}%
    \else
        \hskip -0.5\mylenA\rlap{\usebox\myboxA}{\hskip 0.5\mylenA\usebox\myboxB}%
    \fi}
\makeatother
\makenomenclature

\usepackage{subcaption}
\usepackage{mleftright}
\usepackage{hyperref}

\usepackage{amsmath}
\usepackage{amssymb}
\usepackage{amsfonts}

\usepackage{mdframed}

\newmdenv[leftline=false,rightline=false,linewidth=1pt]{topbot2}

\usepackage{forest}
\usetikzlibrary{external}
\let\oldtikzexternalgetnextfilename\tikzexternalgetnextfilename \renewcommand{\tikzexternalgetnextfilename}[1]{\oldtikzexternalgetnextfilename{#1}\expandafter\tikzsetnextfilename\expandafter{#1}}

\usepackage{pgfplotstable}
\usepackage[outline]{contour}
\contourlength{1.2pt}

\pgfplotsset{compat=1.13} 

\usetikzlibrary{spy}
\usetikzlibrary{calc}
\usetikzlibrary{fadings}
\usetikzlibrary{patterns}
\usetikzlibrary{shadows}
\usetikzlibrary{mindmap}
\usetikzlibrary{backgrounds}
\usetikzlibrary{shapes.symbols}
\usetikzlibrary{shapes.multipart}
\usetikzlibrary{shapes.geometric}
\usetikzlibrary{automata,positioning}
\usetikzlibrary{decorations.fractals} 
\usetikzlibrary{decorations.markings}
\usetikzlibrary{decorations.pathreplacing}
\usetikzlibrary{decorations.pathmorphing}
 
\usepackage{bm}

\usepackage{nomencl}
\makenomenclature
\tikzset{edge from parent/.style={segment angle=10,draw}}

\tikzset{
 my rounded corners/.append style={rounded corners=2pt},
}

\def\BibTeX{{\rm B\kern-.05em{\sc i\kern-.025em b}\kern-.08em
 T\kern-.1667em\lower.7ex\hbox{E}\kern-.125emX}}

\renewcommand{\nomgroup}[1]{%
 \ifthenelse{\equal{#1}{O}}{\item[\textit{Operators}]}{%
 \ifthenelse{\equal{#1}{I}}{\item[\textit{Indices}]}{%
 \ifthenelse{\equal{#1}{A}}{\item[\textit{Acronyms}]}{%
 `\ifthenelse{\equal{#1}{V}}{\item[\textit{Variables and parameters}]}{}}}}}
\usepackage{scalerel}
\usetikzlibrary{svg.path}
\definecolor{orcidlogocol}{HTML}{A6CE39}
\tikzset{
 orcidlogo/.pic={
 \fill[orcidlogocol] svg{M256,128c0,70.7-57.3,128-128,128C57.3,256,0,198.7,0,128C0,57.3,57.3,0,128,0C198.7,0,256,57.3,256,128z};
 \fill[white] svg{M86.3,186.2H70.9V79.1h15.4v48.4V186.2z}
 svg{M108.9,79.1h41.6c39.6,0,57,28.3,57,53.6c0,27.5-21.5,53.6-56.8,53.6h-41.8V79.1z M124.3,172.4h24.5c34.9,0,42.9-26.5,42.9-39.7c0-21.5-13.7-39.7-43.7-39.7h-23.7V172.4z}
 svg{M88.7,56.8c0,5.5-4.5,10.1-10.1,10.1c-5.6,0-10.1-4.6-10.1-10.1c0-5.6,4.5-10.1,10.1-10.1C84.2,46.7,88.7,51.3,88.7,56.8z};
 }
}

\newcommand\orcidicon[1]{\href{https://orcid.org/#1}{\mbox{\scalerel*{ \begin{tikzpicture}[yscale=-1,transform shape]
 \pic{orcidlogo};
 \end{tikzpicture}
 }{|}}}}

\begin{document}

% Title and authors
\title{Grid-Aware Flexibility Operation of Behind-the-Meter Assets: A review of Objectives and Constraints}
\author{Elias~Mandefro~Getie$^{1}$,~\IEEEmembership{Student~Member,~IEEE}~\orcidicon{0000-0001-7776-172X},~{Hossein~Fani$^{1}$,~\IEEEmembership{Student~Member,~IEEE}~\orcidicon{0000-0003-3781-7404},~Md~Umar~Hashmi$^{1}$,~\IEEEmembership{Senior~Member~IEEE}~\orcidicon{0000-0002-0193-6703},~Brida~V.~Mbuwir$^{2}$~\orcidicon{0000-0001-5523-7783},~and~Geert~Deconinck$^{1}$,~\IEEEmembership{Senior~Member,~IEEE}~\orcidicon{0000-0002-2225-3987}}
\thanks{Corresponding author email: eliasmandefro.getie@kuleuven.be}
\thanks{$^{1}$E.M.G, H.F, M.U.H. and G.D are with KU Leuven \& EnergyVille, Leuven/Genk, Belgium}
\thanks{$^{2}$B.M. is with Vito \& EnergyVille, Genk, Belgium}
% \thanks{$^{3}$A.B is with INRIA, DI ENS, Ecole Normale Sup\'erieure, CNRS, PSL Research University, Paris, France.}%
% \thanks{R. D'hulst is with VITO \& EnergyVille, Genk, Belgium}
\thanks{The authors acknowledge the financial support provided by the Flemish Government and Flanders Innovation and  Entrepreneurship (VLAIO) via IMPROcap (HBC.2022.0733) and KU Leuven via BOF scholarship (ZB/23/015).}
% \thanks{979-8-3503-9042-1/24/\$31.00 \copyright~2025 IEEE}
}%

%\author{\IEEEauthorblockN{1\textsuperscript{st} Elias Mandefro Getie}
%%%%%%%%%%%%%%%%\IEEEauthorblockA{\textit{dept. of Electrical Engineering} \\
%%%%%%%%%%%%%%%\textit{KU Leuven/EnergyVille}\\
%%%%%%%%%%%%%%Leuven/Genk, Belgium \\
%%%%%%%%%%%%%eliasmandefro.getie@kuleuven.be}
%%%%%%%%%%%%\and
%%%%%%%%%%%\IEEEauthorblockN{2\textsuperscript{nd} Hossein Fani}
%%%%%%%%%%\IEEEauthorblockA{\textit{dept. of Electrical Engineering} \\
%%%%%%%%%\textit{KU Leuven/EnergyVille}\\
%%%%%%%%Leuven/Genk, Belgium \\
%%%%%%%hossein.fani@kuleuven.be}
%%%%%%\and
%%%%%\IEEEauthorblockN{3\textsuperscript{rd} Md Umar Hashmi}
%%%%\IEEEauthorblockA{\textit{dept. of Electrical Engineering} \\
%%\%%%textit{KU Leuven/EnergyVille}\\
%Leuven/Genk, Belgium \\
%%%%mdumar.hashmi@kuleuven.be}
%%%\and
%%\IEEEauthorblockN{4\textsuperscript{th}Brida Mbuwir}
%\IEEEauthorblockA{\textit{Water and Energy Transition Unit}
%%%%%%%%%\textit{Vito/EnergyVille}\\
%%%%%%%%Genk, Belgium \\
%%%%%%%brida.mbuwir@vito.be}
%%%%%%\and
%%%%%\IEEEauthorblockN{5\textsuperscript{th} Geert Deconinck}
%%%%\IEEEauthorblockA{\textit{dept. of Electrical Engineering} \\
%%%\textit{KU Leuven/EnergyVille}\\
%%Leuven/Genk, Belgium \\
%geert.deconinck@kuleuven.be}

\IEEEoverridecommandlockouts
\IEEEpubid{\makebox[\columnwidth]{979-8-3503-9042-1/24/\$31.00 \copyright 2025 IEEE \hfill } 
\hspace{\columnsep}\makebox[\columnwidth]{\hfill }}

% Make the title area
\maketitle
% Abstract
\begin{abstract} 
 The high penetration of distributed energy resources (DERs) in low-voltage distribution networks (LVDNs) often leads to network instability and congestion. Discovering the flexibility potential of behind-the-meter (BTM) assets offers a promising solution to these challenges, providing benefits for both prosumers and grid operators. This review focuses on the objectives and constraints 
associated with the operation of BTM flexibility resources
% for the operation of behind-the-meter (BTM) resource flexibility 
in LVDNs.  We propose a new classification framework for network-aware flexibility modelling that incorporates prosumer objectives, flexibility sources, and both local and grid-level constraints. This review identifies research gaps in prosumer-centric grid considerations, control strategies, flexibility preferences, and scenarios in the use of BTM resources.
%This review focuses on the objectives and constraints
%associated with the operation of behind-the-meter (BTM)flexibility
%resources in low-voltage distribution networks (LVDN). The high
%penetration of distributed energy resources (DERs) in LVDNs leads
%to network instability and congestion. However, harnessing
%the flexibility potential of BTM assets for both prosumer benefits
%and grid support offers a promising approach to mitigate these
%problems.% Developing network-aware flexibility operations starts with identifying the key objectives of prosumers alongside grid constraints. Thus, considering grid requirements, local objectives such as user comfort and energy savings need to be modelled. This review identifies the objectives of the prosumers, network constraints, control frameworks, flexibility preferences and scenarios in using BTM assets. Finally, the paper highlights the research gaps for further investigation to improve the operation of grid-aware flexibility using BTM assets.
\end{abstract}
% Keywords

\vspace{20pt}
{\textbf{Disclaimer}: This paper is a preprint of a paper submitted to and to be presented at the IEEE ISGT Europe 2025. 
The final version of the paper will be available at IEEE xplore.}
% If accepted, the copy of record will be available at IET Digital Library}

\pagebreak

\tableofcontents

\pagebreak

% Main content of the paper
\section{Introduction}
 In Europe, the deployment of rooftop solar photovoltaics (PV) and the use of electric vehicles (EVs) as one of the targets for EU 2050 net-zero emission is getting more attention . With the massive integration of distributed energy resources (DERs) into the low-voltage distribution network (LVDN), the global energy perspective is experiencing a radical change. However, distribution networks face significant challenges as a result of high penetrations of DER, becoming overloaded and unable to handle dynamic changes in load demand\cite{powerlimit}. Traditional solutions, such as infrastructure reinforcement, usually require significant investments, which are often expensive. 
% due to demand peaks and the unpredictability of variable energy sources. 
Thus,
an alternative promising option is to harness the flexibility potential of BTM DERs and flexible loads. This approach enables distribution systems operators (DSO) to reduce grid congestion by controlling local energy generation and consumption. On the other hand, the transition from consumers to active prosumers in the LVDN presents new challenges for DSOs. Prosumers prioritize getting the maximum benefit from their local generation and maintaining comfort without worrying about the grid. This trade-off between grid requirements and prosumer objectives makes it clear that grid-aware BTM flexibility operation is essential.  

The preferences and capabilities of the prosumers in using BTM DERs affect the objectives of the grid. The uncertainty associated with BTM DERs and user preference requires accurate representation and control frameworks to be exploited for the benefit of the prosumer and the grid \cite{collaborative}. The objectives of individual prosumers are affected by the uncertainty in BTM assets and pose a challenge to the grid operators in monitoring the overall grid stability.  However, considering the grid from the prosumer side based on their flexibility preferences for using BTM assets and the type of flexible loads can help the DSO maintain the operational limits of the network.

\begin{figure}[htbp!]
    \centering
    \includegraphics[width=0.75\textwidth]{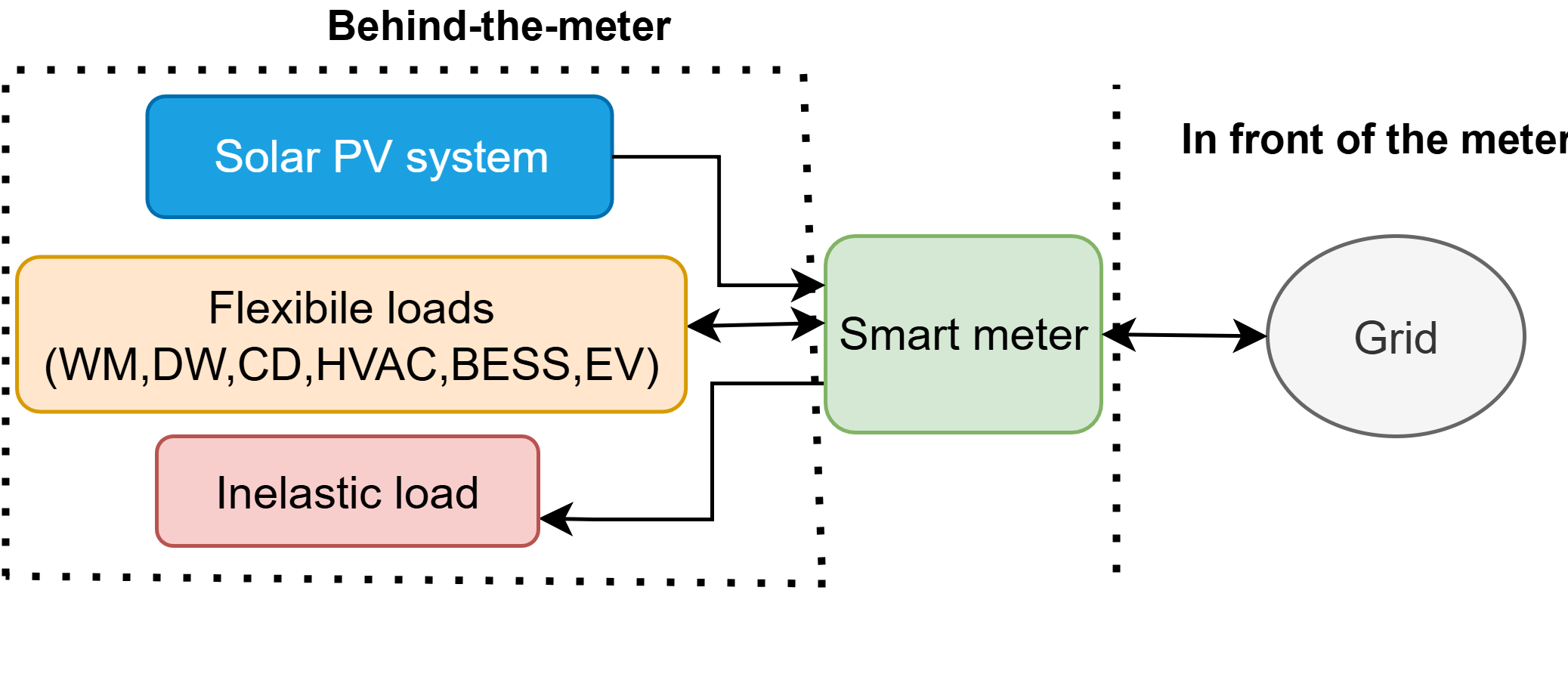} % Replace with your image filename
    % \vspace{-12pt}
    \caption{Descriptive diagram of BTM assets.}
    \label{fig:fig1}
\end{figure}

This review provides a new contribution by identifying the key objectives of residential prosumers, basic local and grid-level constraints, and flexibility preferences that shape grid-aware BTM  flexibility operations.More specifically, it presents a defined classification of network awareness modelling and integrates the flexibility of BTM assets with grid constraints, enabling a comprehensive consideration of both local and grid objectives. Finally, the review highlights research gaps that need further investigation to improve grid-aware flexibility operations of BTM assets in LVDNs.\\
The key contributions of this paper include:
\begin{itemize}
    \item Identifying key prosumer objectives and constraints for grid-aware BTM flexibility operations, including grid constraints. 
\end{itemize}
 \begin{itemize}
     \item Proposing a new classification framework for consideration of the grid in the operation of  BTM asset flexibility. 
 \end{itemize}
\begin{itemize}
    \item Exploring flexibility preferences in using BTM assets. 
\end{itemize}
\begin{itemize}
    \item Identifying research gaps for future exploration.
\end{itemize}
Fig. \ref{fig:fig1} illustrates the BTM assets in residential prosumers, including rooftop solar PV generation, battery energy storage systems (BESS) and flexible loads, such as washing machines (WM), dishwashers (DW), EVs, HVAC, clothes dryer (CD),   heat pumps etc.
% and fixed loads (lights and refrigerators).

The paper is organized as follows: Section \ref{sec2} presents objectives and constraints. Sec. \ref{sec3} network consideration in BTM asset flexibility operation. Sec. \ref{sec4} modelling flexibility preferences in energy systems, focused on BTM assets and flexibility preferences. %Sec. \ref{sec5} discusses control framework and uncertainty modelling. 
Sec. \ref{sec5} discusses the research gaps in the grid-aware flexibility operation of BTM assets. Sec. \ref{sec6} concludes the paper.

\pagebreak
% \vspace{}

\section{Summarizing Objectives and Constraints }
\label{sec2}

\begin{figure}[htbp!]
    \centering
    \includegraphics[width=0.89\textwidth]{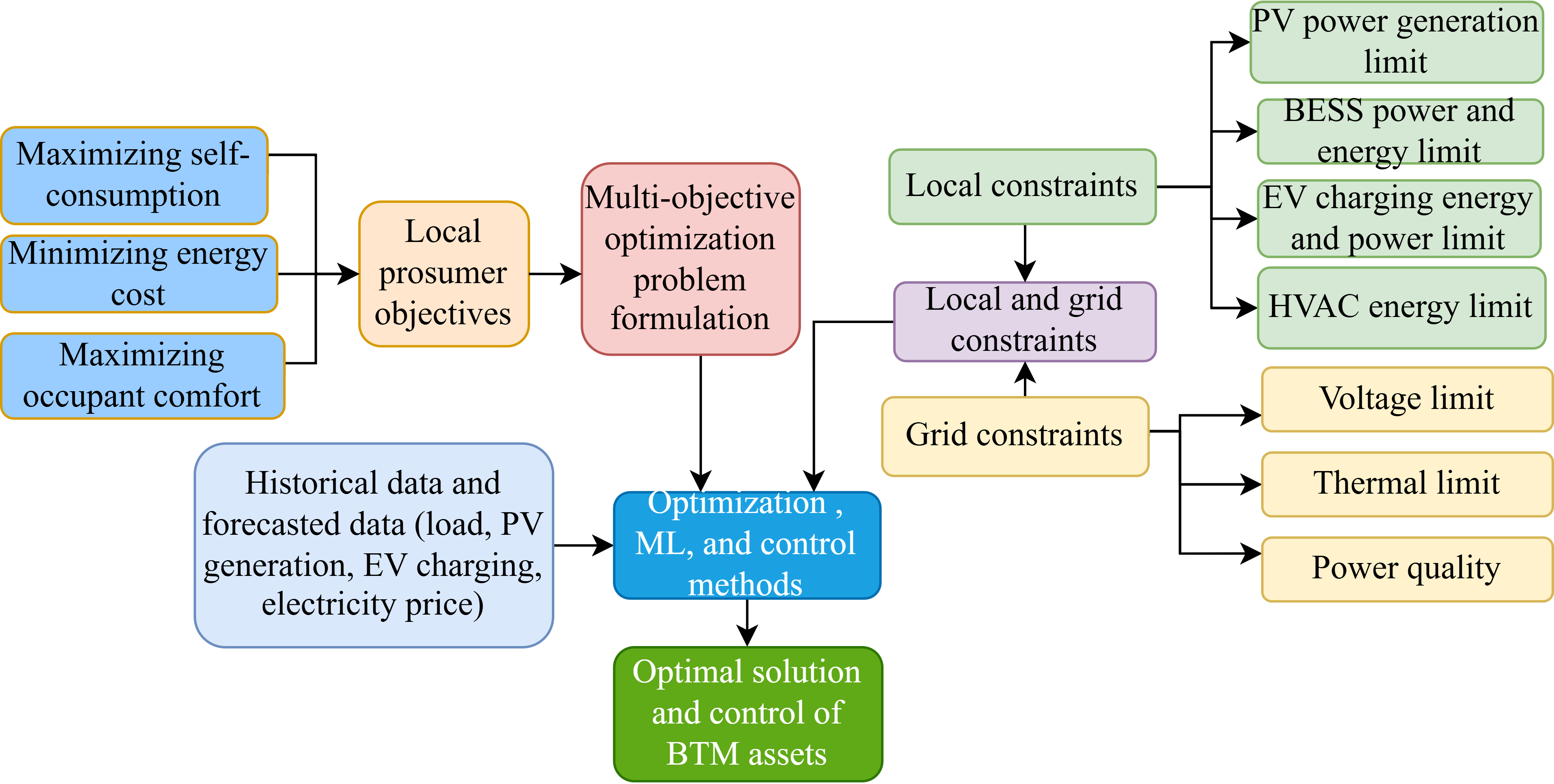} % Replace with your image filename
    % % \vspace{-19pt}
    \caption{{Summary of objectives and constraints of grid-aware BTM assets.}}
    \label{fig:schematic}
\end{figure}

There are two primary actors in unlocking flexibility in LVDNs: prosumers and the DSO. Other entities such as market operators, aggregators and energy retailers are not considered.
Figure \ref{fig:schematic} shows a summary of the objectives and constraints of BTM asset utilization for flexibility operation. The social welfare objectives, such as increasing the hosting capacity (HC) of the grid, achieving greenhouse gas emission targets, ensuring grid stability and fair energy distribution can be addressed using flexibility operation of BTM assets. The increasing HC and grid stability objectives are directly related to the grid constraints such as voltage limits, network capacity limits and power balance. By operating with grid-constraint limits, BTM assets can increase HC and maintain grid stability. The use of BTM assets can support both a local level and the wider social welfare objectives by matching with the operational and grid constraints. In this paper, we focus on prosumer objectives,
local and grid constraints, so no detailed explanation of social
welfare objectives are included. %Next, we detail the objectives and constraints for the grid and flexibility sources. 
% % \vspace{-10pt}

% \vspace{-19pt}

\subsection{Grid operators}
The high integration of DERs in LVDN is leading to an increased chance for grid congestion problems for DSO.%\cite{onmulti}.
On the other hand, BTM DERs can be used to offer grid services such as voltage regulation, congestion management, and peak shaving. The grid operator aims to reduce voltage fluctuation, improve power quality, minimize network congestion, and increase hosting capacity by maintaining the restrictions of the power network\cite{hashmi2025hosting}. The objectives of the grid operator can be integrated into the flexibility operation of BTM assets through technical limits, including power exchange limits, voltage range violations, power balance requirements, peak demand limits, and maximum thermal load capacity \cite{hashmi2023flexibility}. The operational flexibility of BTM assets can serve for local grid congestion management for DSO. Thus, BTM assets can provide flexibility services and support the network operator's objectives by considering the preferences and capabilities of end-users.

\subsection{Flexibility sources of prosumers}
The aim of the prosumer is to use the BTM assets in order to obtain the maximum benefit (example, minimizing cost of consumption etc.) and comfort within the operational limits of the device. In \cite{powerlimit}, the prosumer objectives of minimizing energy cost, maximizing self-consumption, and minimizing the power deviation from the reference power set by the DSO were formulated and implemented using model predictive control (MPC). The authors considered as constraints the EV charging limit, HVAC power consumption limit, and PV generation curtailment. The grid was considered indirectly by setting the peak demand limit and reducing the deviation from the set point reference power and the actual power. This multi-objective optimization problem was formed by using weighting factors to adjust the unit difference and give priority to the prosumer comfort through the decision variables in HVAC power and PV curtailment. While in \cite{uncertain} the objectives of the prosumer were minimizing energy cost and discomfort. The discomfort involved in shifting loads is generated by the flexibility preferences when to activate specific BTM DERs, and the discomfort will increase as moving away from the activation time. The optimization problem is implemented on three levels. The local level scheduling aims to minimize energy costs and discomfort penalty by considering shiftable and interruptible loads. Using the local level scheduling results, the aggregator coordinates the buying and selling of energy between prosumers at the second level, with the driving economic benefits of buying energy from peers being cheaper than from the grid. The last level considers the consumption profile of all prosumers, shiftable and interruptible loads, and as well as electricity tariffs, peak demand, and user preferences to implement flexibility allocation with the same objective as the local level, but with incentives from the DSO by ensuring that peak loads do not exceed the grid capacity.

In \cite{aware} a hierarchical approach was used to coordinate DERs at the prosumer, aggregator, and utility levels to ensure operational flexibility and grid stability. At the prosumer level, the objective focuses on minimizing energy costs and maintaining user comfort. The home energy management system (HEMS) controls the use of BESS, PV,  and HVAC by balancing the energy cost savings with penalties for the deviation from the user’s preferred temperature settings. At the aggregator level, the objective is to minimize grid operating costs and voltage deviations. This includes optimizing the active or reactive power dispatch from the aggregated DERs, reducing curtailment, and compensating participants for providing flexibility and grid services. %The grid was considered using voltage limits at point of common coupling (PCC) and the voltage was calculated by applying a linearized power flow model. 
In \cite{chance} the energy management system targets minimizing the total energy cost while maintaining thermal comfort and adherence to net-zero energy (nZE) sustainability target. The problem was formulated as mixed-integer linear programming (MILP) by incorporating DER-specific constraints and probabilistic constraints for thermal comfort. The optimisation problem was modelled using a chance-constrained MPC approach. The framework controls the HVAC, BESS, EV and PV systems of each building to minimize the overall energy cost and maintain comfort. The grid was considered indirectly by limiting power exchange between the building and the grid to avoid under/over consumption. Ref \cite{energies} works on minimizing the daily energy exchange between the grid and interactive buildings that have BESS and building-integrated photovoltaic (BIPV) systems. The problem was formulated as MILP and modelled using MPC for building energy management by considering the DER capacity limit constraints. In \cite{dynamic} the minimization of the total system cost including energy cost, and curtailment penalties is formulated as MILP for a grid-tied microgrid with PV, EV, BESS and curtailable loads. Load curtailment is considered in the constraint to minimize the energy cost during peak periods. The network is considered by limiting the power at the point of common coupling (PCC). 
 From a prosumer perspective, the following objectives are summarized based on literature reviews in \cite{powerlimit}, \cite{uncertain}, \cite{arbitrage},\cite{networkaware}, \cite{demandside}.
\begin{itemize}
    \item \textit{Minimizing energy costs}: Applying flexibility to BTM assets (PV, EV, BESS, HVAC, WM, heat pump, DW, Dryer), prosumers aim to reduce their energy costs by maximizing their self-consumption and reducing their power import from the grid during peak demand prices.
\end{itemize}
\begin{itemize}
    \item \textit{Maximizing self-consumption}: Prosumers aim to use as much of their locally generated electricity as possible.
\end{itemize}
\begin{itemize}
    \item \textit{Maximizing occupant comfort} (thermal comfort): Prosumers manage BTM assets, focusing on HVAC systems to maintain an optimal indoor environment with a given temperature range.
\end{itemize}

The optimization problem can be solved using various optimization, machine learning (ML) and control techniques by incorporating grid and local constraints. Balancing prosumer and grid objectives requires addressing both local and grid-level constraints, ensuring that the individual prosumer objectives and the operational needs of the grid are consistent with achieving overall social welfare. Both  MPC and reinforcement learning (RL) methods can handle multi-objective optimization problems mostly in a distributed and decentralized (no external communication as shown in Tab. 1 in \cite{hashmi2023robust}) control framework. MPC and RL are the most common control techniques used for the flexibility operation of DERs in LVDN. MPC is commonly used to handle multi-objective optimization problems \cite{powerlimit},\cite{uncertain},\cite{distributed}. It is well-suited for constraint management due to its ability to solve the optimization problem in a rolling-horizon (RH) fashion, continuously changing the control action based on new inputs. The control action can be implemented in centralized, decentralized, and distributed frameworks, while RL is often used in decentralized and distributed control systems \cite{multiagent}, \cite{distributed}. 
%A hybrid method combining MPC and RL leverages the strengths of both approaches to optimize the flexibility operation of BTM assets in LVDN.
The conflicting objectives of minimizing energy costs and maximizing comfort were investigated in \cite{distributed} in a distributed control framework using MPC. The focus was on the thermal comfort of the occupants by controlling the HVAC power consumption. %In this model, comfort related to EV charging preferences and PV curtailment in buildings was not considered.
In \cite{demandside} RL was applied in a decentralized control framework to minimize the energy cost and discomfort by considering the local constraints, grid constraints and uncertainty of DERs. The multi-objective optimization problem was formulated using the weighting sum method, which has the problem of balancing contradictory objectives. In \cite{comparison} maximizing self-consumption of BTM resources and minimizing energy cost for prosumers was studied, considering local constraints in a centralized and distributed control architecture. Table \ref{tab:grid_btm_flex} provides a summary of the existing literature on the operation of BTM asset flexibility. It focuses on methods, objectives, control frameworks, uncertainty modelling, types of flexibility sources/loads, flexibility scenarios, performance metrics,  and the consideration of both local and grid constraints.

\begin{landscape}
\begin{table*}[htbp!]
    \caption{{Summary of Literature on Grid-Aware BTM Flexibility Operation}}
    \centering
    \small
    \renewcommand{\arraystretch}{3.5}
    \resizebox{1.7\textwidth}{!}{
    \begin{tabular}{llccclclcclllll}
    %{|l|l|c|c|c|l|c|l|c|c|l|l|l|l|l|}
        \hline
        \textbf{Ref} & \textbf{Method} & \textbf{\begin{tabular}[c]{@{}c@{}}Maximize\\ self consumption\end{tabular}} & \textbf{\begin{tabular}[c]{@{}c@{}}Minimize\\ energy costs\end{tabular}} & \textbf{\begin{tabular}[c]{@{}c@{}}Maximize\\  comfort\end{tabular}} & \textbf{\begin{tabular}[c]{@{}l@{}}Flexibility \\sources/loads\end{tabular}} & \textbf{\begin{tabular}[c]{@{}c@{}}Local\\ constraints\end{tabular}} 
        & \textbf{\begin{tabular}[c]{@{}l@{}}Control\\ framework\end{tabular}} & \textbf{\begin{tabular}[c]{@{}c@{}}Uncertainty\\ modelling\end{tabular}} & \textbf{\begin{tabular}[c]{@{}c@{}}Grid-\\ constraint\end{tabular}} & \textbf{\begin{tabular}[c]{@{}c@{}}Multi-objective \\ optimization\end{tabular}} & \textbf{\begin{tabular}[c]{@{}l@{}}How is the grid \\ considered?\end{tabular}} & \textbf{\begin{tabular}[c]{@{}l@{}}Flexibility Scenarios\end{tabular}} & \textbf{\begin{tabular}[c]{@{}l@{}}Results/Performance\\ Metrics\end{tabular}} & \textbf{Advantages}\\ \hline %& \textbf{Drawbacks} 
        \cite{powerlimit} & MPC & $\checkmark$ & $\checkmark$ & $\checkmark$ (HVAC) & EV, PV, HVAC & $\checkmark$ & Distributed & $\times$ & $\checkmark$ & $\checkmark$ & Peak demand & Peak shaving (EV, HVAC) &  Peak reduction, cost saving & Peak reduction\\ \hline  %& Limited uncertainty modelling \\ \hline
        \cite{collaborative} & MPC & $\times$ & $\checkmark$ & $\times$ & BESS, DER & $\checkmark$ & Hierarchical & $\checkmark$ & $\checkmark$ & $\checkmark$ & Power balance & Power balance (BESS) &  Reduced grid imports (25\%) & Power balance\\ \hline % & High computational complexity \\ \hline
        %\cite{energies} & MPC & $\times$ & $\checkmark$ & $\times$ & BESS, BIPV & $\checkmark$ & Centralized & $\times$ & $\checkmark$ & $\times$ & Power balance & Power balance(BESS) & Reduced grid exchange by 30\%  & Scalable for small networks \\ \hline %& Centralized, less flexible \\ \hline       
        \cite{chance} & MPC & $\times$ & $\checkmark$ & $\times$ & HVAC, BESS, EV, PV & $\checkmark$ & Centralized & $\checkmark$ & $\checkmark$ & $\times$ & Power limit & Coordination (EV, PV, HVAC)  &  Cost savings (10\%) & Handles uncertainty \\ \hline %& Centralized, less scalable \\ \hline
       \cite{distributed} & MPC & $\times$ & $\checkmark$ & $\checkmark$ (HVAC) & Flexible load & $\checkmark$ & Distributed & $\times$ & $\times$ & $\checkmark$ & $\times$ &  Load shifting (HVAC) &  Cost saving (10\%), comfort & Balance (comfort, cost)\\ \hline %& No grid awareness and uncertainty modelling  \\ \hline
        \cite{multiagent} & RL & $\times$ & $\checkmark$ & $\checkmark$ (HVAC) & PV, HVAC, BESS & $\checkmark$ & Decentralized & $\checkmark$ & $\checkmark$ & $\checkmark$ & Voltage limit & Voltage regulation ( HVAC) & Voltage profile (0.95--1.05 p.u) & Adapts to uncertainty \\ \hline %  Training phase delays 
        \cite{comparison} & RL & $\checkmark$ & $\checkmark$ & $\times$ & PV, SHW storage & $\checkmark$ & Distributed & $\times$ & $\times$ & $\checkmark$ & $\times$ & Self-consumption (PV) & %Cost saving(7\%),
        Self-consumption (25\%) & Flexible control \\ \hline % No grid awareness and uncertainty modelling \\ \hline
         \cite{resilent} & MPC & $\times$ & $\checkmark$ & $\times$ & DER & $\checkmark$ & Hierarchical & $\times$ & $\checkmark$ & $\times$ & Voltage limit & Voltage regulation (DER) & Voltage profile (0.95--1.05 p.u) & Voltage control \\ \hline 
        \cite{policy} & RL & $\times$ & $\checkmark$ & $\checkmark$ (HVAC) & HVAC & $\checkmark$ & Distributed & $\checkmark$ & $\times$ & $\checkmark$ & $\times$ & Thermal comfort (HVAC) &  Cost savings (13.2\%) & Comfort maintained \\ \hline % No grid awareness 
        \end{tabular}
    }
    \label{tab:grid_btm_flex}
\end{table*}
\end{landscape}

\pagebreak

\section{Network consideration modelling}
\label{sec3}
Network consideration in the BTM asset flexibility operation is critical to ensure grid stability and reliability. In energy systems, network-aware considerations refer to the integration of the operational state and constraints of the network into the energy management decision processes \cite{markets}. These considerations can be categorized as direct and indirect, depending on the level of interaction and control with the grid as depicted in Figure \ref{fig:networkCon}. The indirect network consideration can be categorized as active and passive network consideration, whereas the direct grid consideration can be active and implemented in centralized, decentralized, and distributed control frameworks. Active direct methods use live grid data to dynamically control energy flows and provide real-time responses to grid status, whereas passive network consideration is mostly focused on predefined parameters and rules. Indirect grid consideration includes the application of strategies such as demand response to reduce power consumption during peak demand.
\subsection{Indirect network consideration}
From the perspective of prosumers, the network can be considered indirectly, either passively or actively, to meet the operational limit set by the DSO and avoid violating the requirements of the network. The voltage sensitivity factor can be used to measure the voltage at the point of common coupling and check the impact of the BTM DERs on the network. This is an active form of indirect network considerations. The DSO also sets power reference limits to be kept by the prosumers for normal operation of the network. The peak demand management can affect the prosumer and reduce the use of electricity during peak demand periods which indirectly reduces congestion in the network. Prosumers may curtail their solar PV generation to reduce network overloading, which is an indirect way of being network aware. These indirect approaches, without direct control or communication with the DSO, allow prosumers to contribute to grid stability by using their BTM assets.
\subsection{Direct network consideration}
The direct network consideration explicitly considers the distribution network. Examples of direct network consideration could be direct load control, %\cite{lu2012design},
centralized dispatch of flexibility \cite{hashmi2023robustV2} etc.
% In reality, there is no direct network consideration from the prosumer perspective.
The DSO takes a direct role in monitoring the network and flexibility operation of BTM assets. The DSO can apply a centralized control framework, such as direct load control to manage  BTM assets and mitigate network congestion or voltage violation problems. BTM assets can be controlled in decentralized or distributed control, allowing prosumers to participate in network management by considering grid constraints directly. This approach requires coordination between DSO and prosumers. The active involvement of DSO enables BTM assets to operate in a way that supports grid stability.
\begin{figure}[htbp!]
    \centering
    \includegraphics[width=0.97\textwidth]{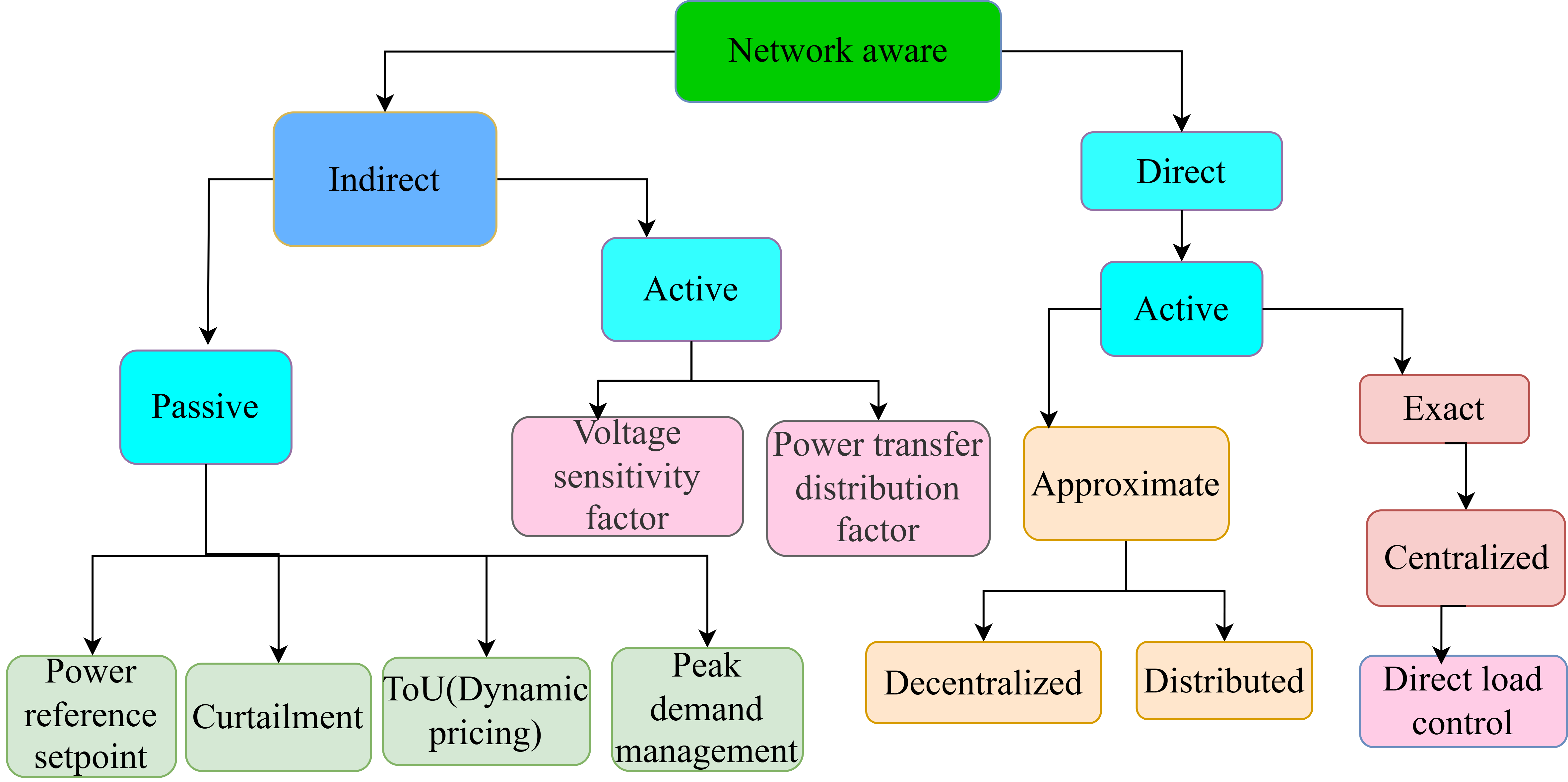} % Replace with your image filename
    % % \vspace{-18pt}
    \caption{Network consideration classification.}
    \label{fig:networkCon}
\end{figure}
Table \ref{tab:network_modeling} classifies the literature based on the framework given in Figure \ref{fig:networkCon}, categorizing the network considerations into direct (active) and indirect (active and passive) by including specific network parameters.
\begin{table}[htbp!]
    \caption{Network constraint consideration }
    % \vspace{-7pt}
    \centering
    \scriptsize % Reduce font size
    \renewcommand{\arraystretch}{1.5} % Adjust row height
    \resizebox{\columnwidth}{!}{ % Use \columnwidth instead of \textwidth
        \begin{tabular}{p{1.6cm}|p{2cm}|p{2.5cm}|p{3.5cm}}
            \hline
            \textbf{Ref} & \textbf{Network Model} & \textbf{Active} & \textbf{Passive} \\ \hline
            % Add your table rows here
        %\cite{distributed,comparison,policy} &$\times$ & $\times$ &$\times$ \\\hline
        \cite{powerlimit},\cite{chance},\cite{dynamic} &indirectly& $\times$ & $\checkmark$peak demand, power limit \\\hline
        \cite{collaborative,energies} &directly&$\checkmark$ power balance&$\times$\\\hline
        %\cite{aware}&directly &$\checkmark$linearized DistFlow model& $\times$\\\hline
     %   \cite{doe} &directly & $\checkmark$voltage limit & $\times$ \\\hline

        \cite{aware,multiagent} &directly&$\checkmark$ voltage limit &$\times$\\\hline
         \cite{btm} & indirectly & $\checkmark$  voltage sensitivity &$\checkmark$ peak demand charge \\\hline
        \end{tabular}
    }
    \label{tab:network_modeling}
\end{table}

\pagebreak

\section{Flexibility Preference Modelling}
\label{sec4}
The preferences of the prosumer are central to model the flexibility operation of behind-the-meter DERs. For EVs, users may have specific mobility patterns (arrival and departure times), charging capacity status by departure time, charging/discharging preferences, and limit the number of charging/discharging cycles \cite{EVPV}. For solar PV systems, users may prefer to increase self-consumption, sell excess energy, and sometimes can be forced to curtail their PV generation due to grid constraints. For HVAC systems, users may have specific comfort ranges and temperature set points and have the flexibility to operate HVAC during low-cost periods. Prosumers can use the BESS for energy arbitrage and backup power as preferences for flexibility modelling. In the flexibility operation of BTM, these preferences are incorporated into the flexibility modelling as a constraint based on the objective of the prosumer. Tabel \ref{tab:flexibility_modeling} summarizes the flexibility preferences and the flexibility modelling applied to specific BTM assets in the literature.

\begin{landscape}
\begin{table*}[htbp!]
    \caption{Flexibility preferences and BTM assets}
    % \vspace{-6pt}
    \centering
    \small
    % \scriptsize % Reduce font size
    \renewcommand{\arraystretch}{1.5} % Adjust row height
    \resizebox{\textwidth}{!}{
    \begin{tabular}{p{1cm}|p{2.5cm}|p{3.35cm}|p{3.45cm}|p{2.0cm}|p{3.45cm}|p{3.15cm}}
        \hline
        \textbf{Ref} & \textbf{Flexibility preferences} & \textbf{EV} & \textbf{Thermostatic loads} & \textbf{PV} & \textbf{BESS} & \textbf{Decision variables}\\ \hline
         \cite{powerlimit} & Self-consumption & EV power limit, SoC limit & HVAC  control, energy limit & PV curtailment & $\times$ & PV , EV  , grid , HVAC power\\ \hline
        \cite{multiagent, collaborative} & Thermal comfort, energy cost & $\times$ & Thermal dynamics, indoor temperature [Tmin, Tmax], air flow rate & PV curtailment  & Battery  power limit,  energy balance & PV power, HVAC power, energy storage capacity\\ \hline
        \cite{chance} & Comfort (thermal, EV) & Energy balance, SoC & Thermal dynamics, temperature  & PV power forecast & Power limit, energy balance & EV , BESS, PV,and HP power\\ \hline
        \cite{dynamic} & EV SoC at departure & Ramp limit, desired EV Soc & $\times$ & PV curtailment & SoC dynamics, battery capacity  & PV, EV, BESS and grid power\\ \hline
       %\cite{collaborative} & comfort  preferences & $\times$ & Linearized HP model & PV curtailment considering power factor limit of inverter & Charge/ discharge power and energy limit, energy balance & HP power, active power load/ generation, comfort parameter\\ \hline
        \cite{EVPV} & EV SoC at departure, time (departure, arrival) & EV power limit, SoC constraint  &$\times$& PV power limit & $\times$ & PV, EV, import/export power\\ \hline
        \cite{predictiveev} & Thermal comfort & PEV  power limit, SoC constraint  & thermal dynamics, HP power limit & Sandia PV model& $\times$ & PV, PEV, HP power\\ \hline
    \end{tabular}}
    \label{tab:flexibility_modeling}
\end{table*}
\end{landscape}

Modelling LVDN flexibility preferences, such as comfort levels for end-users, is crucial for accurately quantifying true flexibility in LVDNs. Flexibility preferences represent the constraints and trade-offs that users are willing to accept when participating in demand-side management or grid services. For example, in residential settings, comfort preferences might include maintaining a specific indoor temperature range or ensuring uninterrupted operation of essential appliances. If these preferences are not explicitly modelled, the assumed flexibility of the system may be overstated, leading to unrealistic expectations about the available capacity for grid balancing or other services. This overestimation can result in operational inefficiencies, grid instability, or even failure to meet user expectations, ultimately undermining the effectiveness of DER-based flexibility programs.

Accurately modelling flexibility preferences ensures that the true potential of DER-based flexibility is quantified in a way that aligns with real-world conditions. By incorporating user comfort and other preferences into the modelling process, grid operators and planners can design more realistic and effective flexibility programs. This approach not only enhances grid reliability but also fosters greater user participation by respecting their needs and constraints. Ultimately, considering flexibility preferences is essential for achieving a balanced and sustainable integration of DERs into the energy system, ensuring that the promised benefits of flexibility are realized without compromising user satisfaction or grid stability.

%\section{Control Framework and Uncertainty Modelling}
\label{sec5}

\pagebreak

\section{Discussion and Future work}
\label{sec6}
The literature review has highlighted several strengths in grid-aware BTM flexibility operations. For example, MPC-based methods demonstrate strong capabilities in constraint handling and multi-objective optimization, making them well suited for incorporating grid constraints such as voltage limits and power balance \cite{powerlimit}. On the other hand, RL approaches offer remarkable adaptability to uncertainty, especially in decentralized frameworks, enabling more prosumer-centric flexibility control \cite{multiagent}. Despite these advantages, both approaches have notable limitations. MPC is computationally intensive, which limits its scalability for large-scale LVDNs. In contrast, RL requires extensive training data and may not ensure real-time compliance with network constraints. In addition, indirect grid-aware strategies:such as peak demand management are easier to implement but often fail to effectively address dynamic grid challenges such as voltage violations. While direct grid-aware control strategies are more robust, they require advanced communication and coordination infrastructure, creating barriers to widespread deployment \cite{aware}. These trade-offs underscore the need for hybrid control approaches and improved forecasting methods to increase the effectiveness and scalability of grid-aware BTM flexibility solutions.

Much of the literature focuses primarily on individual prosumer objectives such as minimizing energy cost,  maximizing self-consumption, and maintaining occupant comfort, often with limited or no consideration of grid constraints. Although these objectives are fundamental for prosumers, the lack of incorporating network constraints can lead to suboptimal results for both the grid and the prosumer. In most of the literature, the grid has been considered as the responsibility of the DSO \cite{distributed},\cite{comparison},\cite{policy}. 
However, this kind of approach overlooks the flexibility potential of prosumers and benefit to the DSO by actively considering the grid constraints. By effectively integrating network constraints such as voltage limits, network congestion, and transformer capacity limits into the optimization problem of the prosumer objectives can increase the benefits to the prosumer and the grid. 
 
The flexibility preferences of the prosumer and grid constraints are key points in the flexibility operation of BTM assets to meet the objectives of the prosumer and the grid. The uncertainty associated with BTM assets and flexibility preferences needs to be incorporated into the flexibility operation by considering the grid. Moreover, controlling the flexibility of BTM assets under uncertainty, along with prosumer flexibility preferences, plays a key role in harnessing the flexibility of BTM assets. By aligning prosumer flexibility preferences with the grid constraints and the uncertainties associated with BTM DERs, flexibility operation of BTM assets can increase  hosting capacity, and improve overall social-welfare objectives. 

Based on the literature review targeting objectives and constraints in grid-aware flexibility operation of BTM assets, the following research gaps are identified:
\begin{itemize}
    \item There is a trade-off in prosumer preferences and grid requirements, focusing on the need for grid-aware flexibility models that balance the prosumer objectives (energy cost, comfort) and grid constraints.  
   % \item Accurate and real-time forecasting methods, including machine learning and stochastic modelling, are essential to improve the performance of grid-aware  flexibility operation of BTM assets. Integrating these forecasting techniques into flexibility operation is still an open challenge.
   % \item Developing grid-aware prosumer centric-models that incorporate the prosumer flexibility preferences and grid constraints into the flexibility operation of BTM assets. 
    \item Developing a privacy-friendly control framework to harness the flexibility potential of BTM assets, both at the individual prosumer level and across LVDN, remains a challenge.
    \item Coordinating the flexibility of multiple BTM DERs across multiple prosumers is complex and requires distributed control and optimization algorithms.
    %\item The socio-economic impact of flexibility operation of BTM DERs requires additional research to assess prosumer preferences, grid investments, equitable benefit access and to quantify the benefits of flexibility. 
    \end{itemize}

Addressing identified research gaps will improve the operation of BTM asset flexibility and mitigate network congestion.

\pagebreak

\section{Conclusion}
\label{sec7}
This paper presents a comprehensive analysis of grid-aware flexibility operations of BTM assets, focusing on prosumer objectives, flexibility preferences, and local and grid constraints. It provides a classification of direct and indirect grid considerations and a comparative analysis of control methods. Among these, MPC and RL are highlighted as key control methods of BTM DERs, which are applied in centralized, decentralized, and distributed control frameworks. The insights from this study contribute to the advancement of BTM DER flexibility operation strategies in LVDN.

%   \section*{Acknowledgments}
% The authors acknowledge the financial support provided by the Flemish Government and Flanders Innovation and  Entrepreneurship(VLAIO) via IMPROcap(HBC.2022.0733) and KU Leuven via BOF scholarship (ZB/23/015).

\pagebreak

% References
\bibliographystyle{IEEEtran}
\bibliography{reference.bib} % Replace with your .bib file

\end{document}